\documentclass[sigconf,authorversion,nonacm,screen]{acmart}

\usepackage{CJKutf8}

\AtBeginDocument{%
  \providecommand\BibTeX{{%
    \normalfont B\kern-0.5em{\scshape i\kern-0.25em b}\kern-0.8em\TeX}}}

\copyrightyear{2024}
\acmYear{2024}
\setcopyright{rightsretained}
\acmConference[CHI EA '24]{Extended Abstracts of the CHI Conference on Human Factors in Computing Systems}{May 11--16, 2024}{Honolulu, HI, USA}
\acmBooktitle{Extended Abstracts of the CHI Conference on Human Factors in Computing Systems (CHI EA '24), May 11--16, 2024, Honolulu, HI, USA}
\acmDOI{10.1145/3613905.3651099}
\acmISBN{979-8-4007-0331-7/24/05}

\usepackage{soul}

\begin{document}
\begin{CJK}{UTF8}{ipxm}

\title[Deceptive, Disruptive, No Big Deal]{Deceptive, Disruptive, No Big Deal: Japanese People React to Simulated Dark Commercial Patterns}

\author{Katie Seaborn}
\email{seaborn.k.aa@m.titech.ac.jp}
\orcid{0000-0002-7812-9096}
\affiliation{%
  \institution{Tokyo Institute of Technology}
  \city{Tokyo}
  \country{Japan}
}

\author{Tatsuya Itagaki}
\orcid{0009-0006-0455-5611}
\affiliation{%
  \institution{Tokyo Institute of Technology}
  \city{Tokyo}
  \country{Japan}
}
\email{itagaki.t.ad@m.titech.ac.jp}

\author{Mizuki Watanabe}
\orcid{0000-0003-3036-7041}
\affiliation{%
  \institution{Tokyo Institute of Technology}
  \city{Tokyo}
  \country{Japan}
}
\email{watanabe.m.ca@m.titech.ac.jp}

\author{Yijia Wang}
\email{wang.y.cf@m.titech.ac.jp}
\orcid{0009-0004-2250-9163}
\affiliation{%
  \institution{Tokyo Institute of Technology}
  \city{Tokyo}
  \country{Japan}
}

\author{Ping Geng}
\email{geng.p.aa@m.titech.ac.jp}
\orcid{0009-0001-8227-7497}
\affiliation{%
  \institution{Tokyo Institute of Technology}
  \city{Tokyo}
  \country{Japan}
}

\author{Takao Fujii}
\email{fujii.t.av@m.titech.ac.jp}
\orcid{0009-0004-5059-3323}
\affiliation{
  \institution{Tokyo Institute of Technology}
  \city{Tokyo}
  \country{Japan}
}

\author{Yuto Mandai}
\email{mandai.y.aa@m.titech.ac.jp}
\orcid{0009-0002-5866-155X}
\affiliation{%
  \institution{Tokyo Institute of Technology}
  \city{Tokyo}
  \country{Japan}
}

\author{Miu Kojima}
\email{kojima.m.ap@m.titech.ac.jp}
\orcid{0009-0006-6122-6750}
\affiliation{%
  \institution{Tokyo Institute of Technology}
  \city{Tokyo}
  \country{Japan}
}

\author{Suzuka Yoshida}
\orcid{0009-0004-8694-3073}
\email{yoshida.s.av@m.titech.ac.jp}
\affiliation{%
  \institution{Tokyo Institute of Technology}
  \city{Tokyo}
  \country{Japan}
}

\renewcommand{\shortauthors}{Seaborn et al.}

\begin{abstract}
Dark patterns and deceptive designs (DPs) are user interface elements that trick people into taking actions that benefit the purveyor. Such designs are widely deployed, with special varieties found in certain nations like Japan that can be traced to global power hierarchies and the local socio-linguistic context of use. In this breaking work, we report on the first user study involving Japanese people (n=30) experiencing a mock shopping website injected with simulated DPs. We found that Alphabet Soup and Misleading Reference Pricing were the most deceptive and least noticeable. Social Proofs, Sneaking in Items, and Untranslation were the least deceptive but Untranslation prevented most from cancelling their account. Mood significantly worsened after experiencing the website. We contribute the first empirical findings on a Japanese consumer base alongside a scalable approach to evaluating user attitudes, perceptions, and behaviours towards DPs in an interactive context. We urge for more human participant research and ideally collaborations with industry to assess real designs in the wild.
\end{abstract}

\ccsdesc[500]{Human-centered computing~Empirical studies in HCI}
\ccsdesc[500]{Human-centered computing~User studies}
\ccsdesc[300]{Human-centered computing~User models}
\ccsdesc[300]{Information systems~Web applications}
\ccsdesc[300]{Information systems~Web interfaces}
\ccsdesc[300]{Human-centered computing~Web-based interaction}
\ccsdesc[500]{Human-centered computing~User interface design}
\ccsdesc[300]{Applied computing~Psychology}

\keywords{Dark Patterns, Deceptive Design, Deceptive Design Pattern, Manipulative Design, Persuasive Design, User Interface Design, User Study, Japan}

\begin{teaserfigure}
  \includegraphics[width=1\textwidth]{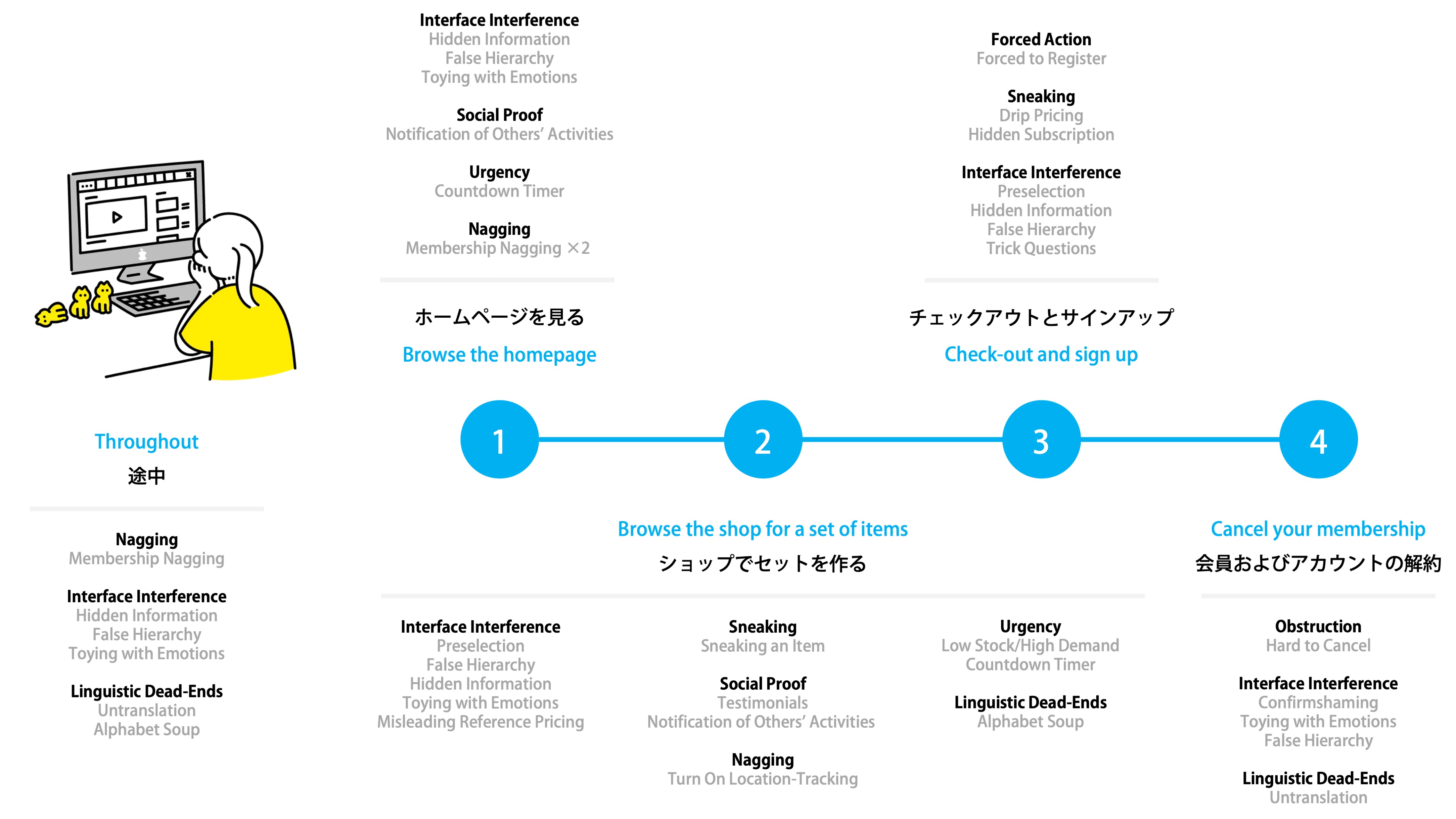}
  \caption{User flow of the website representing a typical online shopping experience involving all DPs. Illustration © \href{https://www.shigureni.com}{shigureni}.}
  \Description{A six-part user flow including browsing the homepage, browsing for a set of items, checking out and signing up, and canceling membership. Also includes DPs experienced throughout the website.}
  \label{fig:teaser}
\end{teaserfigure}


\maketitle

\section{Introduction and Background}

Dark patterns and deceptive designs (DPs) are aspects of user interfaces (UI) that have been specifically designed to manipulate user actions and choices~\cite{gray2023dpsysreview,Brignull_2023}.
Much work has gone into describing~\cite{gray2018darkside,gray2021legal}, finding~\cite{digeronimo2020,hidaka2023linguistic,soe2020norway,gunawan2021webmobile,mathur2019atscale,kyi2023gdpr,yada2022dark}, and regulating~\cite{oecd2022,oecd2023} DPs. However, relatively less work has explored the \emph{user side} of deceptive UIs: how everyday consumers, end-users, and netizens feel about the DPs that they encounter. Notably, few have explored real digital offerings using participant research methods~\cite{borberg2022so,nazarov2022clustering,chaudhary2022videostream,bhoot2021enduser,gray2021edu}. This gap can be traced to the sensitive and even controversial nature of DPs. When portrayed as ``deceptive,'' the implication for companies and other purveyors who employ DPs is negative. Also, whether a particular interface pattern is ``deceptive'' and for whom remains a matter of debate~\cite{fansher2018hashtag,mathur2021whatdark,gray2018darkside,gray2021legal,feng2023analysis,kyi2023gdpr}.

While ideal, the sensitivity of the topic, ethical restrictions, and practical issues have hampered work with real users on real products. Some researchers have not even named the systems evaluated, e.g., \cite{borberg2022so,hidaka2023linguistic,nazarov2022clustering}. One alternative is the use of realistic \emph{simulations}, which include simulated DPs injected into real systems~\cite{roffarello2022steal,utz20219gdprconsent,berens2022cookie,bermejo2021cookie}, imagined designs of complete apps~\cite{bongardblanchy2023,schafer2023countermeasures}, and fully-fledged systems created from scratch~\cite{voigt2021dark,van2022shopping,cranor2022cookie,grassl2021dark,habib2022cookie}. Still, much of this work focuses on specific contexts or DPs in isolation, e.g., app permission requests~\cite{bongardblanchy2023}, cookie consent banners~\cite{cranor2022cookie,bermejo2021cookie,berens2022cookie,habib2022cookie}, the Sneak into Basket/Toying with Emotion DP classes~\cite{van2022shopping}. Moreover, the simulations are often \emph{non-interactive}---screen shots, mockups, and videos, e.g., ~\cite{bongardblanchy2023,schafer2023countermeasures,digeronimo2020}---which limits ecological validity in terms of use case realism and assessment of DPs that require interaction~\cite{carter2008exiting}. Much more work is needed---with a larger volume of potential and real end-users across an array of digital offerings and varieties of DPs---before consensus can be reached.

One emerging user base to study is Japanese consumers. Deception in commercial UIs has become a key topic in Japan, especially as the number of online shoppers continues to grow. Nearly 53\% of households went online to shop in 2022, a sharp incline from 34.3\% in 2018\footnote{\url{https://www.statista.com/statistics/1182675/japan-online-shopping-penetration-households/}}. Last CHI, \citet{hidaka2023linguistic} reported on a heuristic analysis of DPs in the Japanese mobile app market. They found that 
93.5\% of the top 200 Google Play Store apps contained an average of 3.9 DPs~\cite{hidaka2023linguistic}. Moreover, they also discovered the presence of special Japanese DPs linked to global power structures, i.e., the influence of foreign corporations and their local presence in Japan, and the Japanese socio-linguistic context~\cite{hidaka2023linguistic}. Even so, to date, no research in Japan has involved human participants directly experiencing DPs. Japan has also started to take legal action on DPs~\cite{oecd2022}. The Consumer Affairs Agency of the Government of Japan also led the 2023 OECD follow-up report on vulnerability and consumer policy~\cite{oecd2023}.
Even so, the degree to which Japanese netizens are aware of DPs, how they feel about their ``deceptive'' properties, and the implications in typical use cases remain unknown.

To this end, we conducted the first human participants user study on DPs with a diversity of Japanese people. We asked two entwined research questions (RQs): \textbf{\emph{RQ1: How deceptive are DPs to the average Japanese person?}} and \textbf{\emph{RQ2: How does the average Japanese person feel about each form of DP?}} Due to ethics restrictions, we developed a simulation (refer to \ref{sec:website}) containing a range of DPs based on those typically found in real e-commerce websites (refer to \autoref{fig:teaser} and \ref{sec:materials} for details). We found that the deceptibility of people and interfaces varied significantly, and mood was negatively affected by the experience. We contribute:
\begin{itemize}
    \item the first empirical findings on how Japanese consumers perceive a variety of DPs, including the Japanese varieties; and
    \item a methodology for conducting interactive human participant research on DPs using a simulated website that can be translated to in-situ studies.
\end{itemize}
We hope to inspire further interactive, participant-based research on DPs in Japan and elsewhere. We also aim to create a sensitivity-free artefact and training tool for engaging consumers and purveyors alike in discussions on deception in design.


\section{Methods}
We conducted a human factors user study~\cite{stanton2017human} featuring a concurrent think-aloud protocol~\cite{van2003retrospective,alshammari2015ask}, questionnaires, and a semi-structured interview~\cite{blandford2016qualitative}. We received ethics approval (\#2023265) and conducted the study in December 2023.

\subsection{Participants}
We recruited 30 Japanese people with the aid of NHK, a public broadcaster and third-party recruiter, aiming to capture a diverse sample from the larger population. Fourteen women and sixteen men (none of another gender identity) participated. Two were aged 18--24, three 25--34, two 35--44, eight 45--54, nine 55--64, four 65--74, and two 75+. Most had gone to university but not completed studies (n=11), while eight had a bachelor's degree, two a graduate degree, one a degree from a preparatory school, and seven the equivalent of a high school diploma; one had not completed high school yet. 
Most did not know about DPs (only four of 28; for two, this data was missed).
The recruiter used several different participant pools, each with a different rate of compensation; hence, participants were compensated in accordance with each pool at $\sim$5000-8000 yen (USD $\sim$\$30-50) each. Participants were aware and gave consent in advance for their anonymous data and screen footage to be used for media broadcast as well as academic research.

\subsection{Procedure}
Participants were asked to evaluate the interactive features of a new e-commerce website. As in prior work, e.g., \cite{bermejo2021cookie}, the true focus on DPs was hidden to avoid priming~\cite{head1988priming}. Participants were informed about the procedure (without mentioning DPs) and provided consent. They then sat before a laptop and filled out the pre-questionnaire (refer to \ref{sec:q}).

Next, participants watched a training video about the concurrent think aloud protocol procedure. A staple of HCI research, this protocol involves participants speaking their thoughts out loud while performing tasks with a system~\cite{van2003retrospective,alshammari2015ask}. In the video, a researcher performed the protocol while reading a cook book. Then, participants practiced with the same cook book for $\sim$2 minutes.

Next, they experienced the shopping website. They were guided by a researcher through a typical shopping user flow (refer to \autoref{fig:teaser} and \ref{sec:tasks}). They were not forced to carry out the tasks in a certain way or order. This meant that not all participants viewed all pages on the website or experienced all DPs. An observer watched and filled in the observation checklist (refer to \ref{sec:checklist}). After the task, participants filled out the post-questionnaire (refer to \ref{sec:q}). 

Participants were then interviewed about their experience (refer to \ref{sec:semi}). They were told about DPs and shown a PDF containing screen shots of each page in the website. They were asked to point out and describe any notable parts. The observer noted whether they noticed any DPs this second time (refer to \ref{sec:checklist}).


Finally, the host researcher revealed the true nature of the study: to evaluate DPs. Participants were asked to keep the true focus of the study on DPs a secret. Each session took $\sim$50-60 minutes.

\subsection{Materials and Activity}
\label{sec:materials}

\subsubsection{Website Simulation}
\label{sec:website}
\begin{figure*}
    \centering
    \includegraphics[width=\textwidth]{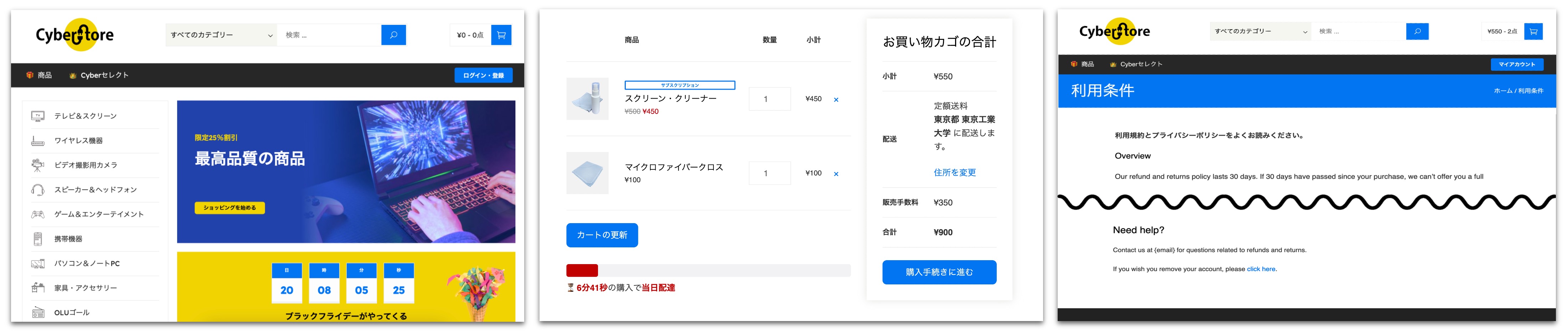}
    \caption{Screen shots of the website: the homepage (left), check-out page (middle), and acceptable use policy page (right).}
    \Description{Three screen shots representing key pages in the ``CyberStore'' website. The first shows the homepage with a leading menu and sale call-outs. The second depicts the check-out page, featuring a sneaked-in product, subtotals, and an urgency timer. The third features the acceptable use policy page, where participants must go to cancel their account.}
    \label{fig:website}
\end{figure*}

A small team of developers created a mock online shopping website for a fake electronics retailer called ``CyberStore,'' (\autoref{fig:website}), which was modelled on e-commerce websites used by Japanese people. A WordPress WooCommerce theme\footnote{Powered by the \href{https://wordpress.org/plugins/woocommerce/}{WooCommerce} \href{https://wordpress.org/}{WordPress} plugin with a customized \href{https://wpcirqle.com/}{WP Cirqle} theme as the base.} was modified to embed DPs within the shopping experience.

\subsubsection{DPs}
\label{sec:dps}
We used the seven categories of DPs compiled in 2022 by the Organisation for Economic Co-operation and Development (OECD)~\cite{oecd2022}, which was based on the taxonomical work of \citet{conti2010malicious}, \citet{bosch2016tales}, \citet{gray2018darkside}, \citet{mathur2019atscale}, and \citet{luguri2021shining}, plus the Japanese-only DPs found in 2023 by \citet{hidaka2023linguistic}: Forced Action, Interface Interference, Nagging, Obstruction, Sneaking, Social Proof, Urgency, and Linguistic Dead-Ends. For each, we created DP \emph{classes}---subtypes of DPs within these categories---and DP \emph{cases}---a point of interaction wherein one or more DPs can be experienced as a gestalt~\cite{digeronimo2020}. These were based on typical examples found in e-commerce websites and sourced from Brignull's ``Hall of Shame.''\footnote{\url{https://www.deceptive.design/hall-of-shame}}
Refer to the visual guide on OSF\footnote{\url{https://osf.io/65wzr}} or the Supplementary Materials for screen shots (Appendix A) and a full list (Appendix C) of each DP case.

\subsubsection{Tasks}
\label{sec:tasks}
\begin{figure*}[!ht]
    \centering
    \includegraphics[width=.985\textwidth]{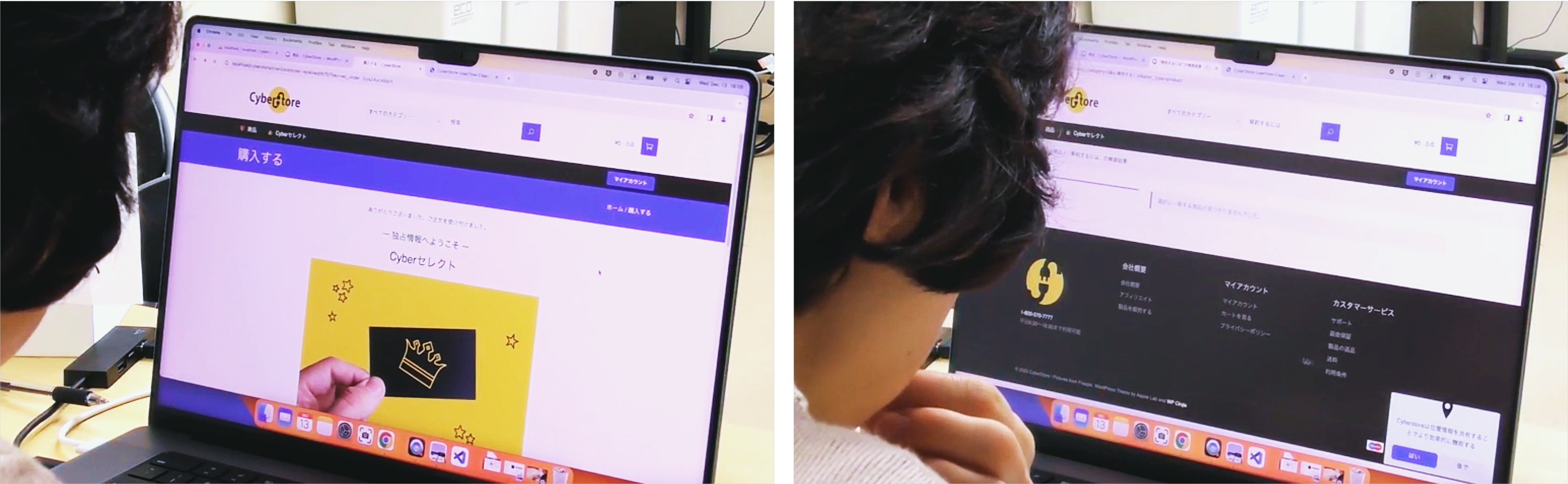}
    \caption{A participant experiences an unintentional premium membership sign-up after making a purchase (left) and an unnecessary location disclosure request pop-up (right).}
    \Description{Two images of a participant interacting with the website. In the first, the participant finds themselves unintentionally signed up for a premium membership. In the second, they consider a pop-up asking for location disclosure.}
    \label{fig:participant}
\end{figure*}

We developed a user flow (\autoref{fig:teaser}) representing typical tasks in an online shopping environment that would also lead participants to experience all DPs within the website (\autoref{fig:participant}). Specifically, the tasks were: (i) Browse the homepage; (ii) Browse the shop for a set of items: a computer, a bag, and cleaning supplies; (iii)  Check-out and sign up; and (iv) Cancel your membership.
Participants were asked to carry out the tasks in this order. However, they were not forced to complete each task nor prevented from exploring the website. In doing so, they were additionally able to experience other DPs placed throughout the website. The DPs that could be experienced per task are outlined in \autoref{fig:teaser}. The tasks were rigorously developed to ensure realism and encounters with all DPs, as well as timed in pilot tests with Japanese lab members. Participants were given a 3-minute limit for the last task.

\subsection{Data Collection}

\subsubsection{Questionnaire Instruments and Measures}
\label{sec:q}
We used pre- and post-task questionnaires. We used valence and arousal to assess affective state before and after the experience. This duo is among the most common subjective measures of emotion, with arousal in particular found to thoroughly explain affective states~\cite{Mauss2009}. For self-reporting, we used the valence and arousal dimensions from the face-based tactile version~\cite{iturregui2020towards} of the Self-Assessment Manikin (SAM)~\cite{bradley1994measuring}. This language-free self-report instrument uses a 9-point semantic differential scale, where 1 is low valence/arousal and 9 is high valence/arousal. In the post-task questionnaire, we used the Japanese translation~\cite{yamano2015jpsus} of the 10-item System Usability Scale (SUS)~\cite{lewis2018system} to assess acceptability~\cite{bangor2008empirical} with a 5-point Likert scale.

\subsubsection{Observation Checklist and Deceptibility Metric}
\label{sec:checklist}
We created a checklist for observers to record whether each participant \emph{noticed} each DP and their \emph{reaction} to it. To account for priming~\cite{head1988priming}, the checklist distinguished between the participant first experiencing the website, after being prompted to comment on the page (as part of the concurrent think-aloud protocol), and the semi-structured interview, which involved revealing the focus on DPs and the user flow (refer to \ref{sec:semi}). The checklist also accounted for the chance that participants were not tricked by certain DPs or found alternative paths through the website, thus missing certain DPs. The \emph{noticed} options were: on their own; after prompting; during the reveal; missed chance; and did not notice. The \emph{reaction} options were: felt manipulated or without a choice, and felt negative (``deceptive''); disturbed or interrupted and wanted to avoid it (``disruptive''); accepted it (``no big deal''); and no reaction, i.e., tricked.

We used the \emph{noticed} data to create a deceptibility metric based on the lead researcher's expertise in DPs and as a designer and scholar with over 15 years experience in UX. We added the ratios of each option with weights: .7 for noticing on their own; .25 for after prompted, and .05 for after learning about DPs. We multiplied this sum by the quotient of 10/7 and inverted the product for
a measure between 0 and 1, where 0 is undeceived and 1 is fully deceived.

\subsubsection{Semi-Structured Interview}
\label{sec:semi}
We used semi-structured interviews~\cite{blandford2016qualitative} with a funnel approach~\cite{wilson2013interview} to avoid priming responses while also capturing prior familiarity with and understanding of DPs. 
The questions were: \emph{Have you heard of ``dark patterns'' before? Have you experienced dark patterns before? Were there any aspects of the design that you felt were deceptive during operation? What other feelings or impressions do you have about the design?}
Refer to Appendix B for the Japanese versions. The participant was shown a PDF (refer Appendix A) of the user flow and asked to point out specific aspects of the design that aligned with their perceptions and feelings. The observer recorded their responses in the checklist.

\subsection{Data Analysis}
We generated descriptive statistics by participant, DP class, DP case, and DP category. We used the acceptability interpretation of the SUS from \citet{bangor2008empirical}: 51.6 as unacceptable, 51.7--71 as marginal, and 71.1--100 as acceptable. We ran exploratory inferential statistics to evaluate affect before and after the experience (paired t-test) and relationships between deceptibility and other variables (parametric and/or non-parametric, depending on the data types and normality results via Shapiro-Wilk tests). One participant did not supply ``before'' affect scores and another did not fill out the SUS; their data was excluded from analyses using these measures. Quotes were sourced from the interviews.


\section{Results}

\subsection{Deception: Ability, Noticeability, Trickery, and Deceptibility (RQ1)}

\subsubsection{Task Completion}
Two DP cases affected the path through the website: (i) a Hidden Subscription to the ``CyberSelect'' premium membership through Trick Wording and Pre``Un''Selection of a checkbox at checkout and (ii) Untranslation of the account cancellation page (Hard to Cancel). For (i), half unintentionally signed up for ``CyberSelect.'' 
Of the 42\% that noticed, 20\% accepted it, 6\% felt it was deception, and 75\% felt it was disruptive. Most (n=27, 90\%) were unable to complete (ii). Of the 35\% that noticed, deceptive (72\%) or disruptive (15\%), but some accepted it (13\%).

\subsubsection{Noticeability and Reactions to Trickery}
Descriptive statistics on the noticeability and reactions to all 25 DP cases are in Supplementary Tables 2 and 3 (refer to Appendix D). Most (58\%) DP cases were unnoticeable (M=17.3, SD=7, MD=19.3, IQR=12). When noticed, 47\% cases were deemed ``no big deal'' (M=6.4, SD=6, MD=5, IQR=3.5), 29\% were deceptive (M=3.7, SD=4.9, MD=2, IQR=2), and 24\% were disruptive (M=2.6, SD=2.9, MD=2, IQR=2). The most noticeable were the homepage call-outs (n=25, 83\%), the low stock indicators (n=24, 80\%), the testimonials on the product pages (n=23, 77\%), and Sneaking an Item (n=23, 77\%). The least noticeable were the CyberSelect membership fee featuring Alphabet Soup and the Social Proof on the search dropdown (only three each).

Descriptive statistics for DP category and class are in Supplementary Table 4 (Appendix D). The most noticeable category was Obstruction (80\%) and the least was Forced Action (73\%). Obstruction was most deceptive (53\%), Interface Interference was most disruptive (68\%), and Urgency was most acceptable (41\%). For classes, Sneaking an Item (Sneaking) and Testimonial (Social Proof) were most noticeable (23\%), while Alphabet Soup (Linguistic Dead-End) and Misleading Reference Pricing (Interface Interference) were least noticeable (90\%).
Sneaking an Item (Sneaking) (58\%), Hard to Cancel (Obstruction) (53\%), and Untranslation (Linguistic Dead-End) (48\%) were most deceptive. Trick Questions (Interface Interference) were most disruptive (53\%). Low Stock, High Demand (Urgency) (67\%) and Testimonials (Social Proof) (63\%) were most acceptable.

\subsubsection{Deceptibility Metric}
We calculated deceptibility metrics for participants and DPs (refer to Appendix D).

Participant deceptibility ran the gamut: a Shapiro-Wilk test did not find a statistically significant departure from normality, \emph{W}(30) = .98, \emph{p} = .783, indicating a bell curve. The highest score was 0.92 and the lowest was 0.40 (M=0.7, SD=0.1, MD=0.7, IQR=0.1). There were no statistically significant correlations to age, gender, or education. This indicates that everyone was susceptible to DPs, erring on the side of being more deceived overall (skewed -0.4).

\begin{figure*}[!ht]
    \centering
    \includegraphics[width=\textwidth]{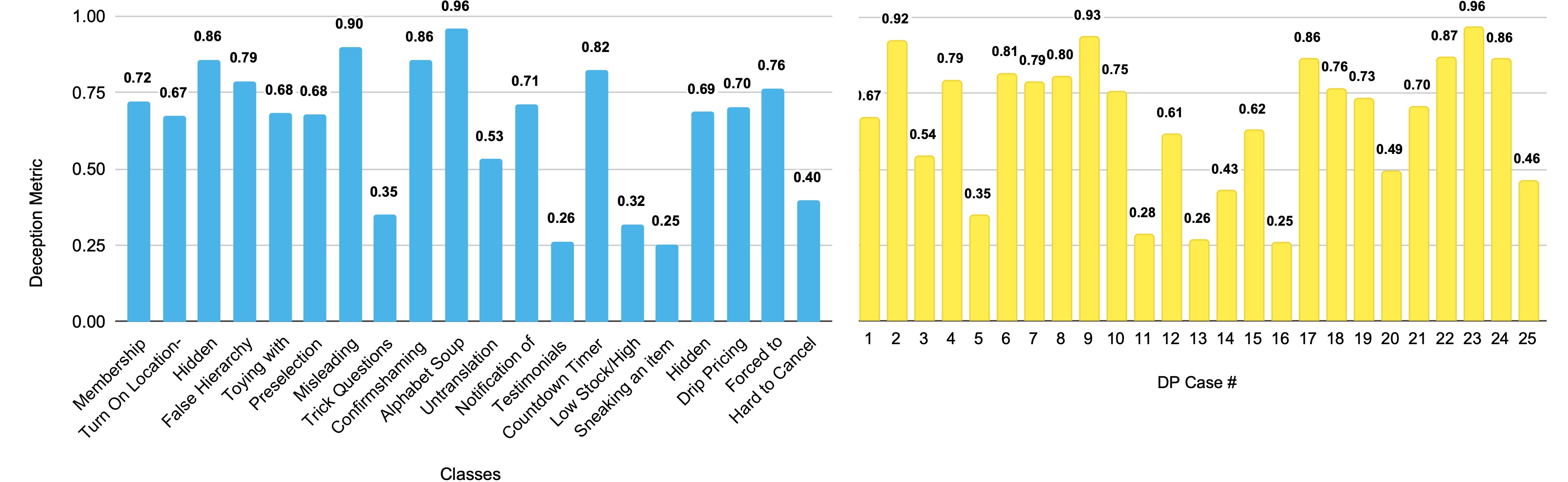}
    \caption{Deceptibility metric scores for classes (left) and cases (right).}
    \Description{Two graphs featuring the deceptibility metric scores for all classes and cases.}
    \label{fig:graphdeception}
\end{figure*}

The deceptibility of DP cases (\autoref{fig:graphdeception}, right) varied (M=0.7, SD=0.2, MD=0.7, IQR=0.3) but were skewed high (-0.6). A Shapiro-Wilk test indicated a statistically significant departure from normality, \emph{W}(25) = .92, \emph{p} = .047. 
The most deceptive were the CyberSelect fee with Alphabet Soup (0.96), search dropdown with Social Proofs and False Hierarchies (0.93), and membership price comparison chart (0.92). The least were sneaking in an item (0.25), the product page testimonials (0.26), and the low stock indicators (0.28).

At a higher level, the deceptibility of DP categories and classes (\autoref{fig:graphdeception}, left) also varied (M=0.6, SD=0.2, MD=0.7, IQR=0.2). A Shapiro-Wilk test found a statistically significant departure from normality, \emph{W}(28) = .92, \emph{p} = .033, skewed high (-0.7). The most deceptive categories were Forced Action (0.76), Interface Interference (0.75) and Linguistic Dead-Ends (0.75). The least was Obstruction (0.40). The most deceptive classes were Alphabet Soup (0.96), Misleading Reference Pricing (0.90),
Hidden Information (0.86) and ConfirmShaming (0.86). The least were
Sneaking an Item (0.25), Testimonials (0.26), Trick Questions (0.35), and Hard to cancel (0.40).

In short, some DPs were more deceptive than others, though most were deceptive. Later, we will discuss the nuances of these results given the noticeability results and reactions, as well as the feelings of participants, up next.

\subsection{Feelings: Mood and Judgments (RQ2)}

The SUS scores indicated that most participants did not accept the website (M=35.5, SD=32.5, MD=32.5, IQR=32.5, hi=85, lo=0). 69\% (n=20) deemed the website unacceptable, with three accepting it and six with marginal scores.

A paired-t test showed a statistically significant large difference between valence before (M=6.9, SD=1.4) and after (M=4.6, SD=2.5), \emph{t}(28) = 5, \emph{p} < .001. Since the data was non-normal, we confirmed this result with a Wilcoxon Signed-Rank test, which indicated the same for before (M=7, n=29) and after (MD=5, n=29), \emph{Z} = -3.7, \emph{p} < .001, \emph{r} = -0.8.
In short, valence decreased significantly for most people right after experiencing the website.
Results of a paired-t test for arousal also indicated a statistically significant large difference before (M=2.9, SD=2) and after (M=5.5, SD=2.4), \emph{t}(28) = 5.5, \emph{p} < .001.
Arousal greatly rose for most people after the experience. Taken together, the valence and arousal results suggest that people became significantly more agitated after the experience.

Quotes from participants help understand these results. As A2 explained, ``I was concerned about being solicited for information. The location request pop-up worried me, now that I think about it, because I pressed it without paying attention to anything.'' The particularly problematic Untranslation DP was summed up by A11: ``I couldn't delete my account. The page was written in English.'' A20 provided further context: ``I feel that the content of the English text may be important, or that I am being deceived.'' B6 helped make sense of why Social Proofs---fake reviews and testimonials by fake people---were generally noticeable and also not considered deceptive: ``The reviews were mostly good ... it's word-of-mouth.'' Sneaking in an Item was among the most noticeable and least deceptive according to the deceptibility matrix, but among those who were not tricked, most (58\%) considered it deceptive. A27 explained: ``That cloth comes with the cleaner. They force you to buy it. They don't mention it in the description.'' We will discuss this variability and the nuanced relationship between noticeability and trickery as represented in the deceptibility metric next.


\section{Discussion}

Dark patterns and deceptive designs are multifarious UI phenomena. In the case of our online shopping simulation, nearly two-thirds went unnoticed (RQ1). For the rest, noticeability and deceptibility varied (RQ1), as did people's reactions and feelings towards each pattern (RQ2). We now attempt to bring a depth of understanding to these findings by considering particular examples.

Almost no one was able to delete their account. This shows the power of the newly found Untranslation DP as a mode of Obstruction. Untranslation is tied to knowledge of the local linguistic context and relies on assumptions of language ability. We used English in our simulation, given its global reach and power, as well as statistics on Japan's low English proficiency~\footnote{\url{https://www.nippon.com/en/japan-data/h01843/}}. The statistics bore out in our case. Other nations with low English proficiency may also be at risk of Untranslation, but this has yet to be explored.

A gestalt of DPs led half of participants to unknowingly sign up for the premium member service. Yet, rather than deceptive, those that realized found it disruptive. This may point to attribution biases~\cite{heider2013psychology} that may be self-serving~\cite{miller1975self}. Whether people blamed themselves or attributed the problem to the system will need to be researched. At least, our participants assumed good intentions, which raises the issue of user vulnerability to such DPs.

Reactions to the Social Proofs raise the issue of subversive influences. Repeated exposure to media stimuli can lead to unconscious shifts in preference by way of familiarity~\cite{cox2002beyond}. Social Proofs, among the most and least noticeable, may be implicated, given the social media age we live within. Most participants deemed these DPs acceptable. This may be explained by social mobilization effects, where even small profile pictures can have massive effects on behaviour, perhaps especially when representing ``friends'' and influencers~\cite{bond201261}. 
The interviews also suggest a perceived benefit of ``word-of-mouth'' information. Japan has been characterized as a collectivist market culture, where social bonds are prioritized and group acceptance of commercial offerings dictate market value~\cite{Synodinos2001}. We must be wary of taking advantage of people's implicit trust in social indicators.
Similarly, Urgency DPs were also considered noticeable and acceptable, perhaps because of the store context and repeated exposure to these rather widespread DPs in the wild.

Perhaps the most subversive DP was Alphabet Soup, found in two membership fee displays. This DP was noticed only by a handful of participants and was a top-scorer on the deceptibility metric (0.96 and 0.92). This study is the first to empirically explore this novel DP and its sibling Untranslation, so it is difficult to contextualize these results within previous research. However, the National Consumer Affairs Center of Japan has recently drawn attention to this DP in the form of the \yen \phantom{} symbol, which is used for both Chinese and Japanese currencies\footnote{\url{https://www.kokusen.go.jp/news/data/n-20230419_2.html}}. Consumers and companies must be cautious of conflating these symbols, as their emergence represents a sublime form of deception for the Asian market.

Misleading Reference Pricing, a form of Interface Interference, was also highly unnoticeable and deceptive, according to our metric. Indeed, virtually no one noticed it, even when shown the user flow and told about DPs. This is a common DP in shopping websites and may escape the notice of the average consumer, allowing it to live on even when laws change in response to advocacy initiatives. We will need to raise attention to this DP and critically engage designers, purveyors, and stakeholders~\cite{gray2021legal} in light of regulatory shifts around less deceptive forms of DPs.

Despite its name, ``Sneaking'' DPs were the least deceptive and easily noticed. Obstruction, too. These may be more ``blunt'' forms of DPs, irritants rather than deceptive. Controlled experimental work with and without these DPs may confirm and provide further context on these findings from a consumer perspective~\cite{voigt2021dark}.

\subsection{Limitations}
We used a simulation that was developed to the level of commercial viability, containing DPs based on real cases from live websites. Nevertheless, we recognize that this limits the ecological validity of our results. Our simulation also contained significantly more DPs than the average 3.9 previously found for apps in the Japanese Google Play Store~\cite{hidaka2023linguistic}. This may have lessened the realism of the experience. At the same time, our results confirm that most DPs went unnoticed. Still, future work should study UIs with more realistic levels of DPs. Importantly, other kinds of UI, such as smartphone apps and video games, as well as other kinds of online stores, such as clothing stores and grocery markets, should be explored, as the level and range of DPs may vary by platform.

We acknowledge limitations related to recruitment. We relied on a third-party recruiter publicly known to participants (NHK). This allowed us to recruit a diverse and random selection of participants from the Japanese populace. However, participants may have self-selected in or out of the study based on the recruiter. We also recruited a relatively small sample (N=30) and did not have an experimental design with a control setup, limiting the generalizability of our findings. Finally, we did not control or capture relevant metrics, such as daily Internet usage, type of degree or career, or knowledge of English (for the Untranslation DP).

Finally, this was a pilot study. Our protocol was not preregistered, and we adapted our procedure and methods based on the pilot tests, notably switching from retrospective to concurrent protocol due to timing issues and adding options to the observer checklist. A preregistered study will be conducted next. We also created a novel metric of deceptibility that could be applied to individuals as well as DPs. This weighted measure was useful for determining relative deceptibility, especially by taking into account priming and exposure to DP stimuli. Still, it rests on our operationalizations of performance and arbitrary weights. This metric should be vetted and discussed with experts in design and DPs, modified, if needed, and then tested empirically. We also did not measure affective reactions to specific DPs; targeted measurement can do so in future.


\section{Conclusion}
We have taken a first step towards better understanding the deceptibility and UX of DPs commonly deployed in online stores for the Japanese context. We have demonstrated variability in perceptions and deceptions while highlighting an overall trend of deceptibility for most Japanese participants and DPs. The next step is experimental work with larger sample sizes and real commercial UI. 

\balance

\begin{acks}
Our sincere gratitude to NHK, especially Tatsuro Imono and Yuya Higashi, for funding part of this research (development and participant fees) and handling the recruitment of participants. We sincerely thank Émilie Fabre for development assistance. We thank \href{https://tokyotech-i.co.jp/}{Tokyo Tech Innovation} and Dai Senoo for supporting this research in various ways. We thank Peter Pennefather for pre-reviewing this manuscript. Katie Seaborn and Takao Fujii conscientiously dissent to in-person participation at CHI this year; read their positionality statement here: \url{https://bit.ly/chi24statement}
\end{acks}

\bibliographystyle{ACM-Reference-Format}
\bibliography{refs}

\end{CJK}
\end{document}